\documentclass[twocolumn,showpacs,showkeys]{revtex4}
\usepackage{graphicx}
\usepackage{bm}
\usepackage{color}
\usepackage{amsmath}
\usepackage{natbib}

\newcommand{\vect}{\boldsymbol}

\begin{document}

\title{On a consistent macroscopic description for a spin quantum plasma with interparticle interactions}

\author{Pavel A. Andreev}
\email{andreevpa@physics.msu.ru}
\affiliation{M. V. Lomonosov Moscow State University, Moscow, Russia.}
\author{Felipe A. Asenjo}
\email{faz@physics.utexas.edu}
\author{Swadesh M. Mahajan}
\email{mahajan@mail.utexas.edu}
\affiliation{Institute for Fusion Studies, The University of Texas at Austin, Texas 78712, USA.}

 \date{\today}

\begin{abstract}

Quantum mechanical averaging of the particle concentration
operator is an effective starting point for derivation of the
many-particle quantum hydrodynamic equations.
In many-particle quantum systems, we have to separate the ordered motion of the local center
of mass (velocity field), the thermal, and the quantum motion. The quantum mechanical average process, invoked here, is completely determined, and is different from the usual averaging processes  that introduces undefined probabilities for quantum states.
It is shown that the Madelung decomposition for the
$N$-particle spinor wave function allows the correct introduction of
the velocity field, and gives explicit expressions for the quantum contributions to
both the momentum, and the spin flux. The formalism also contains
plasma effects produced by the Coulomb and spin-spin interparticle interactions.
It is shown that both interactions appear in plasma dynamics as effective electric and magnetic fields.
As result we find a fully coherent description for spin quantum plasma in the
self-consistent field approximation for interparticle interactions. A simple consequence --
 the change brought about by  interparticle correlations on the propagation of electrostatic waves
 in a spin quantum plasma -- is discussed.

 \end{abstract}

\pacs{52.25.-b, 67.10.-j, 67.30.hj}% PACS, the Physics and Astronomy
                             % Classification Scheme.
\keywords{spin quantum plasma, interparticle correlations}%Use showkeys class option if keyword

\maketitle

%%%%%%%%%%TEXT

\section{Introduction}

When a one-particle Schrodinger equation is  expressed as  a set of two
``hydrodynamical equations": the continuity equation and the Euler equation,
\cite{Haas PRE 00,Haasbook}, it differs from the classical Euler equation in that the
thermal pressure in Euler equation is replaced
by an explicit  quantum pressure  whose origin is the
quantum Bohm potential.  The latter, proportional to the
$\hbar^{2}$, is caused by the nonlocal nature of the particle formed by a de Broglie wave packet.

For scalar particles, only two real classical functions (an amplitude defining the one-particle quantum fluid density, and a phase that characterizes the one-particle velocity as  a potential field) are needed to represent the wave function. This decomposition produces the quantum Bohm potential \cite{Haas PRE 00,Haasbook}. For spin-1/2 particles, obeying Pauli-Schr\"{o}dinger (P-S) equation, however, the spinor  wave functions require a generalization \cite{Marklund PRL 07, pavel}, that give rise to new spin contributions  to quantum pressure, as well as a spin evolution equation.

While the ``fluidization"  of one-particle Schr\"{o}dinger and P-S systems is straightforward, the quantum ensemble average that will lead to
a desirable macroscopic fluid description for a spin quantum plasma has to be carefully performed. In the often quoted works
(see for example Refs.~\cite{Haasbook, Marklund PRL 07}), the ensemble average is understood to be over the normalized occupation probabilities for a given mixture of quantum states. In this form, each macroscopic quantity is defined as the sum over the corresponding fluid-like elements modulated by its probability of occurrence. The macroscopic equations, so obtained, have been investigated for  a host of new interesting  linear and nonlinear quantum-magnetohydrodynamical phenomena (see, for example, Refs.~\cite{brodinMHD, misracircularly,shuklawave,masood,zamanian,zama2,marksoliton,misrasoliton,misra2soliton, markferro,brodinpondero,misrawis,lundin} and references therein) including new instabilities driven by spin \cite{markferro,misramodu,dae,akbari,vitaly,braun}. A new revealing formulation casts the system  into a  vortical  structure where the  the spin-dynamics is shown to resemble the vortical-dynamics of an ideal fluid \cite{Mahajan PRL 11}; extensions to relativistic quantum plasmas have also been constructed \cite{asenjo}.

Although the procedure developed in Refs.~\cite{Haasbook, Marklund PRL 07} may be consistent for non-interacting particles, it is difficult to see how it could deal, properly, with $N$-wavefunctions correlations (as those produced by Pauli exclusion principle), and inter-particle interactions.
In  this work , we seek to develop a more consistent quantum mechanical averaging process to derive many-particle quantum hydrodynamic equations. The system is defined through a many-particle Schrodinger equation with a many-particle wavefunction $\psi(\textbf{r}_{1},...,\textbf{r}_{N},t)$ [shortly $\psi(R,t)$], where $\textbf{r}_{p}$ is the coordinate  where the  $p$-th particle is located;
$R$ designates the  whole coordinates set.
Thus, we are able to introduce Coulomb and spin-spin interparticle interactions constructing a fully coherent description for spin quantum plasma in the
self-consistent field approximation.
One of the tests that the new formalism must pass is that it will produce the corresponding one-particle limit in the approximation of independent particles.
Similar attempts in kinetic theories for quantum plasmas has been proposed \cite{jensC}.

The paper is organized as follows. In Section \ref{sec2}, we present the basis for the macroscopic equations using the quantum averaging for the $N$-particle Pauli equation. In Sec. \ref{sec3} we obtain the fluid equations using the generalized Madelung decomposition for the wave function.
The fluid version of the $N$-particle Pauli equation will contain the contribution of the spin, producing additional forces in the Euler equation and an evolution equation for the magnetic moment. Besides, the momentum flux and the spin flux of both equations contain quantum and thermal effects.
We show that the inter-particle forces produce effective electric and magnetic fields that interact with the fluid velocity, and with the spin fields. In Sec. \ref{sec4} we discuss the approximations needed to obtain a close form for the Bohm potential, and in Sec. \ref{sec5} we give the vortical description for the present formulation. In Sec. \ref{sec6} we study the contribution of the previous mentioned effects for the propagation of electrostatic waves. And in Sec. \ref{sec7} we discuss our results.

\section{Construction of macroscopic equations}
\label{sec2}

The dynamics of the $N$-interacting fermions is given by the Pauli-Schrodinger equation
\begin{equation}
i\hbar\partial_t\Psi(R,t)=\hat H\Psi(R,t),
\label{pseq}
\end{equation}
with the Hamiltonian
\begin{equation}\label{Hamiltonian1}
  \hat H=\sum _{p=1}^N\left(\frac{1}{2m}D_p^i D_p^i+q\phi_{p}-\mu \sigma_p^i B^i_{p}\right)\,+\sum_{p,n,p<n}H_{pn} ,
\end{equation}
acting on the $N$-particle wave function $\Psi(R,t)$. Einstein convention over repeated latin indices will be used throughout this paper. Here $m$ and $q$ are, respectively,  the mass and charge of the particle, $\mu=q\hbar/(2m c)$ is the magnetic moment, $\hbar$ is the reduced Planck constant and $c$ is the speed of light. The covariant derivative operator is
\begin{equation}\label{}
  D_p^i=-i\hbar\partial_p^i-\frac{q}{c} A^i_{p}\, ,
\end{equation}
and $\phi$ and  $A^i$ are the scalar and vector potential, $\sigma^i$ are the Pauli matrices and $B^i$ are the components of the magnetic field. The last term in the Hamiltonian \eqref{Hamiltonian1} describes the long-range interaction between particles in a self-consistent field approximation \cite{MaksimovTMP 2001,pavel}
\begin{equation}\label{}
  H_{pn}=q^2 G_{pn}-\mu^2 G_{pn}^{ij}\sigma_p^i\sigma_n^j\, ,
\end{equation}
 where the first term represents the Coulomb interaction and the second term is the spin-spin interaction. The  two Green functions are:
 \begin{equation}\label{green1}
 G_{pn}\equiv G_{pn}(|\mathbf r_p-\mathbf r_n|)=\frac{1}{|\mathbf r_p-\mathbf r_n|}\, ,
 \end{equation}
and \cite{MaksimovTMP 2001,pavel}
\begin{eqnarray}\label{green2}
 G_{pn}^{ij}&\equiv& G_{pn}^{ij}(|\mathbf r_p-\mathbf r_n|)\nonumber\\
 &=&\partial_p^i\partial_n^j\left(\frac{1}{|\mathbf r_p-\mathbf r_n|}\right)+4\pi\delta^{ij}\delta(\mathbf r_p-\mathbf r_n)\, .
 \end{eqnarray}

The quantum mechanical average  that will yield the macroscopic variables, can be defined in a standard way using the $N$-particle wavefunction.  The microscopic density \cite{MaksimovTMP 1999,pavel,Andreev IJMP 12,Landau Vol 3,Weinberg book,Klimontovich book}, for instance, will take the form,
\begin{equation}\label{density1}
  n\equiv n(\mathbf r, t)=\int d R \sum _{p=1}^N\delta(\mathbf r-\mathbf r_p)\Psi^\dag(R,t) \Psi(R,t)\equiv\langle \Psi^\dag \Psi\rangle\, ,
\end{equation}
with $dR=\Pi_{p=1}^N d\mathbf r_p$. In the rest of the paper, $\langle \xi \rangle\equiv \int d R \sum _{p=1}^N\delta(\mathbf r-\mathbf r_p) \xi$ will  symbolize the average of the quantity $\xi(R,t)$.

By taking the time derivative of ~\eqref{density1}, we derive the first conservation law, the continuity equation:
\begin{equation}\label{}
  \frac{\partial n}{\partial t}+\partial_i J^i=0\, ,
\end{equation}
where the macroscopic current ($h.c.$ stands for hermitian conjugate)
\begin{equation}\label{current}
  J^i\equiv J^i(\mathbf r,t)=\frac{1}{2m}\left\langle\Psi^\dag D^i_p \Psi+h.c.\right\rangle\, ,
\end{equation}
is the appropriately averaged microscopic current. Following Ref.~\cite{pavel}, the evolution equation for the current $J^i$ is obtained by taking the time derivative of \eqref{current} and invoking the P-S equations \eqref{pseq} and \eqref{Hamiltonian1}

\begin{eqnarray}\label{Euler eq gen}
  \frac{\partial J^i}{\partial t}+\frac{1}{m}\partial_j \Pi^{ij}&=&\frac{q}{m}n E^i+\frac{q}{mc} \varepsilon^{ijk} J^j B^k+\frac{1}{m}M^k \partial^iB^k\nonumber\\
  &&-\frac{q^2}{m}\int d\mathbf r'\partial^i G(\mathbf r,\mathbf r') n_2(\mathbf r,\mathbf r',t)\nonumber\\
  &&+\frac{1}{m}\int d\mathbf r' \partial^i G^{jk}(\mathbf r,\mathbf r')M_2^{jk}(\mathbf r,\mathbf r',t)\, ,\nonumber\\
  &&
\end{eqnarray}
where $E^i\equiv E^i(\mathbf r,t)$ and $B^i\equiv B^i(\mathbf r,t)$ are the macroscopic electric and magnetic fields.
%, and the summation for the Green functions \eqref{green1} and \eqref{green2} has been extended to integrals.
Four macroscopic density tensors are introduced in \eqref{Euler eq gen}. The first one is equivalent to the  momentum flux in fluid dynamics,
\begin{equation}\label{flux1}
  \Pi^{ij}\equiv \Pi^{ij}(\mathbf r,t)=\frac{1}{4m}\left\langle \Psi^\dag D^i_p D_p^j\Psi+\left(D_p^i\Psi\right)^\dag D_p^j\Psi+h.c.\right\rangle\, ,
\end{equation}
and the second one is the magnetic moment density
\begin{equation}\label{magnetization}
  M^i\equiv M^i(\mathbf r,t)=\mu\left\langle \Psi^\dag \sigma_p^i \Psi\right\rangle\, .
\end{equation}

The other two new fields are related to inter-particle interactions. The two-particle number density (corelation function)
\begin{equation}\label{2density}
n_2(\mathbf r,\mathbf r',t)=\int d R \sum _{p,n,p<n}\delta(\mathbf r-\mathbf r_p)\delta(\mathbf r'-\mathbf r_n)\Psi^\dag(R,t) \Psi(R,t)\, ,
\end{equation}
is created by Coulomb correlations, and the tensor
\begin{eqnarray}\label{2magnetization}
&&M_2^{ij}(\mathbf r,\mathbf r',t)=\nonumber\\
&&=\mu^2\int d R\sum _{p,n,p<n}\delta(\mathbf r-\mathbf r_p)\delta(\mathbf r'-\mathbf r_n)\Psi^\dag(R,t)\sigma_p^i\sigma_n^j \Psi(R,t)\, ,\nonumber\\
&&
\end{eqnarray}
is the manifestation of the spin-spin interaction between two particles.

In order to formally complete the particle dynamics, we must also calculate the dynamical equation that governs the evolution of the macroscopic spin magnetization. Taking the time derivative of the magnetic moment density, we derive \cite{pavel}
\begin{eqnarray}\label{magn moment evol}
  \frac{\partial M^i}{\partial t}+\partial_j \Lambda^{ij}&=&\frac{2\mu}{\hbar}\varepsilon^{ijk}M^j B^k\nonumber\\
  &&+\frac{2\mu}{\hbar}\varepsilon^{ijk}\int d\mathbf r'G^{km}(\mathbf r, \mathbf r')M_2^{jm}(\mathbf r, \mathbf r',t)\, ,\nonumber\\
  &&
\end{eqnarray}
where the new magnetic moment flux tensor  stands for
\begin{equation}\label{flux2}
  \Lambda^{ij}\equiv\Lambda^{ij}(\mathbf r,t)=\frac{\mu}{2m}\left\langle \Psi^\dag \sigma_p^i D_p^j\Psi+h.c.\right\rangle\, .
\end{equation}

Finally, the whole system must be  coupled to the Maxwell equations
\begin{equation}\label{}
  \partial_i B^i=0\, , \qquad \partial_i E^i=4\pi q n\, ,
\end{equation}
\begin{equation}\label{}
  \varepsilon^{ijk}\partial_j E^k=-\frac{1}{c}\partial_t B^i\, ,
\end{equation}
\begin{equation}
\varepsilon^{ijk}\partial_j B^k=\frac{4\pi q}{c}J^i+4\pi \varepsilon^{ijk}\partial_j M^k+\frac{1}{c}\partial_t E^i\, ,
\end{equation}
where the total current is formed by the charged fluid current $J^i$ and the magnetization current $\varepsilon^{ijk}\partial_j M^k$.

\section{Macroscopic fluid description}
\label{sec3}

We now proceed to construct a
hydrodynamical theory  from the macroscopic equations derived in the last section.
The first step is to  do a Madelung decomposition of the $N$-particle wave function (see, for instance, Ref.~\cite{Holland})
\begin{equation}\label{madelung}
\Psi(R,t)=a(R,t)\exp\left(\frac{i}{\hbar}S(R,t)\right)\varphi(R,t)\, ,
\end{equation}
where $a$ is an amplitude, $S$ is a phase and $\varphi$ is a normalized  spinor ($\varphi^\dag\varphi=1$).

We expect that  the Madelung decomposition, first invoked  for   spinless particles, will help to obtain an explicit form of the quantum Bohm potential \cite{MaksimovTMP 1999,Andreev PRB 11}. We point out  the decomposition used in Refs.~\cite{MaksimovTMP 2001,pavel} for the $N$-particle wavefunction of spinning particles  failed to provide the correct convective part of the momentum $\Pi^{ij}$, and spin $\Lambda^{ij}$ fluxes. We will soon see that the  Madelung decomposition (\ref{madelung}) does, indeed, yield the correct explicit form  for $\Pi^{ij}$ and $\Lambda^{ij}$.

We first, using Eq.~\eqref{density1}, see that the macroscopic density becomes
\begin{equation}\label{}
  n=\left\langle a^2\right\rangle\, .
\end{equation}

The Madelung decomposition implies, for the $p$-th particle, a microscopic velocity
\begin{equation}\label{}
  v_p^i(R,t)=\frac{1}{m}\partial_p^i S -\frac{i\hbar}{m}\varphi^\dag\partial_p^i\varphi-\frac{q}{mc}A^i_{p}\, ,
\end{equation}
that will help write down the macroscopic current \eqref{current} as
\begin{equation}\label{}
  J^i=\left\langle a^2 v_p^i\right\rangle\, ,
\end{equation}
and a macroscopic fluid velocity as
\begin{equation}\label{}
  V^i\equiv V^i(\mathbf r,t)=\frac{J^i(\mathbf r,t)}{n(\mathbf r,t)}\, .
\end{equation}

In the same way, from the microscopic spin
\begin{equation}\label{}
s^i_p(R,t)=\varphi^\dag\sigma_p^i\varphi\, ,
\end{equation}
normalized to unity, we derive the magnetic moment density \eqref{magnetization}
\begin{equation}\label{}
  M^i=\mu\left\langle a^2 s_p^i\right\rangle\, ,
\end{equation}
such that the macroscopic spin density emerges as
\begin{equation}\label{}
  S^i(\mathbf r,t)=\frac{M^i(\mathbf r,t)}{\mu n(\mathbf r,t)}\, .
\end{equation}

Let us now evaluate the momentum flux \eqref{flux1} in terms of the newly defined fluid variables,
\begin{eqnarray}\label{flux3}
  \Pi^{ij}&=&\left\langle \frac{\hbar^{2}}{2m}\left(\partial_p^i a \partial_p^j a-a\partial_p^i\partial_p^j a\right)\right.\nonumber\\
  &&\left.+a^2 m v_p^i v_p^j+\frac{\hbar^2 a^2}{4m} \partial_p^is_p^k \partial_p^j s_p^k\right\rangle\, .
\end{eqnarray}
The first two terms of the tensor create the Bohm potential, the third contributes to the fluid pressure, and the last term is a force produced by the self-interaction of the spin particle. The magnetic moment flux density \eqref{flux2} is, similarly, expressed as
\begin{equation}\label{flux4}
  \Lambda^{ij}=\mu\left\langle a^2 s_p^i v_p^j-\frac{\hbar a^2}{2m}\varepsilon^{imk}s_p^m\partial^j_p s_p^k\right\rangle\, .
\end{equation}
As a further step in the ``hydrodynamic" project, we will introduce the normal as well as the spin temperatures. Let us split
\begin{equation}\label{}
  z_p^i=v_p^i-V^i\, , \qquad w_p^i=s_p^i-S^i\, ,
\end{equation}
such that $\langle a^2 z_p^i\rangle=0$ and $\langle a^2 w_p^i\rangle=0$. The field $z_p^i$ signifies thermal speed, and $w_p^i$ is the displacement (thermal) fluctuation of the spin about the macroscopic average $S^{i}$. Both displacements will be used in the fluxes \eqref{flux3} and \eqref{flux4}. Note that the macroscopic spin is no longer normalized to unity because of the thermal-spin effects,
\begin{equation}\label{macrospinnorm}
  S^i S_i=1-\frac{1}{n}\langle a^2 w^i_pw_p^i\rangle\, .
\end{equation}

Because the evolution of the fluid equations is dominantly driven by the electromagnetic field interaction, we could perform a self-consistent approximation for the correlations \cite{pavel}
\begin{eqnarray}\label{selfapprox}
  n_2(\mathbf r,\mathbf r',t)&\approx & n(\mathbf r,t)n(\mathbf r',t)\, ,\nonumber\\
  M^{ij}_2(\mathbf r,\mathbf r',t)&\approx & M^i(\mathbf r,t)M^j(\mathbf r',t)\, .
\end{eqnarray}

We are ready now  to write down the  dynamical equations for the spin quantum plasma explicitly in term of the fluid variables:  we begin with the continuity
\begin{equation}\label{continuityevolution}
  \frac{\partial n}{\partial t}+\partial_i \left( n V^i\right)=0\, ,
\end{equation}
 the momentum evolution equations
\begin{widetext}
\begin{eqnarray}\label{momentumevolution}
  \left(\frac{\partial }{\partial t}+V^j\partial_j\right) V^{i}&=&\frac{q}{m} {\widehat E}^i+\frac{q}{mc} \varepsilon^{ijk} V^j B^k-\frac{1}{n}\partial_j p^{ij}+\frac{\mu}{m} S^j \partial^i {\widehat B}^j+\frac{\hbar^2}{8 m^2}\partial^i\left(\partial_j S^k\partial_j S^k\right)-\frac{1}{nm}Q^i\, ,\nonumber\\
  &&
\end{eqnarray}
\end{widetext}
where we have used the relation \eqref{ap1}. We have also introduced new definitions:  the pressure tensor $p^{ij}=\langle a^2 z_p^iz_p^j\rangle$,  the effective electric
\begin{equation}\label{eEf}
  {\widehat E}^i=E^i-q\, \partial^i\int d\mathbf r'  G(\mathbf r',\mathbf r)n(\mathbf r',t)\, ,
\end{equation}
and the effective magnetic field
\begin{equation}\label{eBf}
  {\widehat B}^i=B^i+\frac{\hbar c}{2qn}\partial_j\left(n\partial^j S^i\right)+\mu\int d\mathbf r' G^{ij}(\mathbf r',\mathbf r) n(\mathbf r',t) S^j(\mathbf r',t)\, .
\end{equation}
The last term in \eqref{momentumevolution} is the quantum force produced by the Bohm potential and the thermal-spin interactions
\begin{eqnarray}\label{Q curr}
  Q^i&=&\partial_j\left\langle\frac{\hbar^2}{2m}\partial_p^j\left(a\partial_p^i a\right)-\frac{\hbar^2}{m}a\partial_p^i\partial_p^j a+\frac{a^2}{m}\partial_p^j w_p^k\partial_p^i s_p^k \right.\nonumber\\
  &&\left.\qquad+\frac{a^2}{m}\partial_p^j s_p^k\partial_p^i w_p^k -\frac{a^2}{m}\partial_p^i w_p^k\partial_p^j w_p^k  \right\rangle\nonumber\\
  &&-\frac{\hbar^2 n}{8m}\partial^i\left[\frac{1}{n}\partial_j\left[n\partial^j\left(\frac{1}{n}\left\langle a^2 w_p^k w_{p}^k \right\rangle\right)\right]\right]\, ,
\end{eqnarray}
and it is rather complicated.

In Eq.~\eqref{momentumevolution},  the effective electric field \eqref{eEf} contains the intrinsic electric field generated by Coulomb interaction, which is, naturally, the gradient of a potential. The effective magnetic field \eqref{eBf} is also enhanced by the magnetic field generated by the variations of the spin field, and by the intrinsic magnetic field produced by interparticle spin-spin magnetization. In this sense, the present effective magnetic field is a quantum generalization of the one used in Ref.~\cite{Mahajan PRL 11}. The fifth term on the right-hand side (rhs) is a perfect gradient and acts as an additional fluid pressure originating in the self-interaction of spin fluctuations.

Finally, the spin evolution equation takes the form
\begin{equation}\label{macrospineq}
  \left(\frac{\partial}{\partial t}+V^j\partial_j\right) S^{i}=\frac{2\mu}{\hbar}\varepsilon^{ijk}S^j {\widehat B}^k-\frac{1}{n}{\cal Q}^i\, ,
\end{equation}
where
\begin{equation}\label{Q spin}
  {\cal Q}^i=\partial_j\left\langle a^2 z_p^j s_p^i-\frac{a^2}{m}\varepsilon^{imk}s^m_p\partial_p^j w_p^k\right\rangle\, ,
\end{equation}
stands for all the thermal-spin interactions. To derive \eqref{macrospineq}, we used  \eqref{ap2}.
From Eq.~\eqref{macrospineq} we note that the macroscopic spin is normalized only if the quantum spin-thermal interaction ${\cal Q}$ is neglected. This is consistent with \eqref{macrospinnorm}.

\section{Approximate form of quantum Bohm potential}
\label{sec4}

Due to many-particle effects, the  force $Q^i$ ~\eqref{Q curr} appearing in Eq.~\eqref{momentumevolution} \cite{MaksimovTMP 1999,Andreev PRB 11} is rather complicated, and  contains implicit averages that defy an expression in terms of hydrodynamic variables, the particle concentration, the velocity field and the average spin field.  In order to feel confident that the quantum pressure, though, somewhat opaque, is really correct, we  would show that in the simplest approximation (noninteracting particles), it reduces to the correct known limit of the quantum pressure in one-particle QHD, the Bohm potential.

We will concentrate on the first two terms of $Q^i$, which can be extracted out  of the averaging operator in the approximation of weakly interacting particles. This transforms the quantum force in  \eqref{Q curr} to
\begin{eqnarray}\label{Q curr2}
  Q^i&\approx & -\frac{\hbar^2n}{2m}\partial^i\left(\frac{\triangle\sqrt{n}}{\sqrt{n}}\right)\nonumber\\
  &&+\frac{1}{m}\partial_j\left\langle{a^2}\left(\partial_p^j w_p^k\partial_p^i s_p^k +\partial_p^j s_p^k\partial_p^i w_p^k -\partial_p^i w_p^k\partial_p^j w_p^k\right) \right\rangle\nonumber\\
  &&-\frac{\hbar^2 n}{8m}\partial^i\left[\frac{1}{n}\partial_j\left[n\partial^j\left(\frac{1}{n}\left\langle a^2 w_p^k w_{p}^k \right\rangle\right)\right]\right]\, .
\end{eqnarray}
that explicitly contains the correct Bohm potential in the first term on the rhs.

\section{Vortical structure}
\label{sec5}

If thermal-spin interactions are neglected, the quantum force \eqref{Q curr2} becomes a perfect gradient of the Bohm potential, then the system of equations \eqref{continuityevolution}, \eqref{momentumevolution} and \eqref{macrospineq} are reducible to a  vortical structure that looks exactly like the one developed in \cite{Mahajan PRL 11}. The principal result of Ref.~\cite{Mahajan PRL 11}, dwelling on the standard formulation of  the Pauli-Schrodinger  equation, is the construction  of  a classical--quantum hybrid vorticity
\begin{equation}
\Omega_-^i=\Omega_c^i-\frac{\hbar c}{2q}\Omega_q^i\, ,
\end{equation}
from the classical $\Omega_c^i=B^i+(mc/q)\varepsilon^{ijk}\partial_j V_k$, and the quantum part $\Omega_q^i=(1/2)\varepsilon^{ijk}\varepsilon^{lmn}S_l\partial_jS_m\partial_kS_n$ \cite{refvort}. The hybrid vorticity obeys the Helmholtz vortical dynamics leading to a new conserved  hybrid-helicity
\begin{equation}
h_{-}=\int d^3 x\, \Omega_-^i P_{-}^{i}\, ,
\end{equation}
such that $d h_{-}/dt=0$. Here, $P_{-}^{i}$ is the generalized potential defined by $\Omega_-^i=\varepsilon^{ijk}\partial_jP_{-}^k$. Notice that ${\widehat B}^i$ given in \eqref{eBf} contains the extra term (the Green function integral representing correlations), and thus generalizes all previous results including those of Ref.~\cite{Mahajan PRL 11}. For more details on the spin quantum vortex dynamics, we refer the reader to \cite{Mahajan PRL 11}.

\section{Electrostatic waves}
\label{sec6}

The interparticle  Coloumb and spin-spin interactions, cause, {\it inter alia}, interesting modifications to classical wave propagation. As an illustration, we will calculate the electrostatic waves in a two-dimensional plasma with isotropic pressure $p$. In this case, the Green function \eqref{green2} has been shown to be \cite{pavelproc}
\begin{equation}\label{greenss2}
  G^{ij}=-\frac{\delta^{ij}-3 l^i l^j/l^2}{l^3}+\frac{8\pi}{3}\delta^{ij}\delta(\vect l)\, ,
\end{equation}
where $l=|\vect l|\equiv |\mathbf r_p-\mathbf r_n|$, and $l^i$ stands for the $i$-component of $\vect l$.

Let us explore a simple model in which the  two-dimensional plasma is composed of electrons ($q=-e$; $\mu=-e\hbar/2mc<0$) moving in a plane perpendicular to the external constant magnetic field $B_0 \hat e_z$ and with $p^{ij}=\delta^{ij} p/m$. To study electrostatic waves, the total fluid density fluid is taken to be $n=n_0+\delta n$, where $n_0$ is the background density and $\delta n \ll n_0$ is the perturbation. In the absence of an equilibrium flow, the perturbed  fluid velocity, $\delta V^i$, is such that $\delta V^z=0$. The perturbations of the macroscopic spin field $\delta S^i$ are perpendicular to the normalized equilibrium spin field, that  is aligned antiparallel to the external magnetic field to minimize the magnetic moment energy. All perturbed quantities vary as $\exp(i k_j r^j-i\omega t)$, where $r^i$ and $k^i$ are restricted to the $x-y$ plane. The perturbed continuity equation simplifies to
\begin{equation}\label{per1}
  \delta n=\frac{n_0 k_i\, \delta V^i}{\omega}\, ,
\end{equation}
while (neglecting the thermal-spin interactions) the linearized momentum equation yields
\begin{eqnarray}\label{per2}
  -i \omega \delta V^i&=&-\frac{e}{m}\delta E^i-\omega_c\, \varepsilon^{ijz}\delta V_j\nonumber\\
  &&-i\frac{k^i\delta n}{k^2 n_0}\left(v_s^2 k^2+\frac{\hbar^2 k^4}{4 m^2}+\frac{\mu^2 k^3 n_0}{m}\beta(k)\right)\, ,\nonumber\\
  &&
\end{eqnarray}%%+\frac{2\pi q^2 n_0 k}{m}
where $\delta E^i$is the electric field, $\omega_c= e B_0/mc$ is the cyclotron frequency and $v_s=\sqrt{k_B T/m}$ is the thermal speed.
For simplicity, we consider electric field perturbations through Poisson equation (however to include the intrinsic electric field in \eqref{eEf} is straightforward).
The last term, representing the effect of correlations, contains the function
\begin{equation}\label{47}
  \beta(k)=2\pi \int_{k\, l_{min}}^{\infty} \frac{dx}{x^2}J_0(x)\, ,
\end{equation}
where $J_0$ is the Bessel function, of order 0 and $l_{min}$ is the minimum distance for the spin-spin interaction. This integral appears from the spin-spin interaction  in the definition \eqref{eBf} for the effective magnetic field  (see Appendix \ref{apendiceB}). It must be evaluated numerically.

In addition, the perturbed spin evolution equation becomes
\begin{eqnarray}\label{per3}
  -i\omega \delta S^i&=&\left(\frac{2\mu B_0}{\hbar}+\frac{2\mu^2 n_0}{\hbar}\chi- \frac{\hbar k^2}{2m}\right) \varepsilon^{ijz}\delta S_j\nonumber\\
  &&+\frac{2\mu^2 n_0}{\hbar}\varepsilon^{ijz} \Upsilon_{jk}\,  \delta S^k
\end{eqnarray}
with
\begin{equation}\label{49}
  \chi=\int_{r_{min}}^\infty d \mathbf r' \frac{1}{|\mathbf r'-\mathbf r|^3}=2\pi\int_{l_{{min}}}^\infty \frac{d l}{l^2}=\frac{2\pi}{l_{min}}\, ,
\end{equation}
and
\begin{eqnarray}
 \Upsilon_{j k}\equiv \int d\mathbf r' G_{j k}(\mathbf r',\mathbf r) e^{i \mathbf k\cdot(\mathbf r'-\mathbf r)}\, ,
\end{eqnarray}
which is evaluated in \eqref{b4green2spin}.

Finally, the perturbed Poisson equation for a two-dimensional plasma reads \cite{krash}
\begin{equation}\label{per4}
  i k_i\, \delta E^i=-2\pi e k \delta n\, .
\end{equation}

\subsection{Spin waves}

Because the modes are electrostatic, the spin equations \eqref{per3} are decoupled from the rest. Consequently there are two independent  branches of the dispersion relation that can be labelled, respectively, the spin waves and the plasma waves. The spin wave dispersion relation
\begin{equation}
\label{spindisp1}
 \omega^2= Q\left(Q-\frac{4\pi\mu^2 n_0 k}{\hbar}\right)
 \end{equation}
 where
\begin{equation}\label{spindisp2}
 Q= -\frac{2\mu B_0}{\hbar}+\frac{\hbar k^2}{2m}-\frac{4\pi\mu^2 n_0}{\hbar\, l_{min}}\, ,
\end{equation}
can be further simplified in the ``long" wavelength limit $2\pi k\, l_{min} \ll 1$, i.e, when $4\pi\mu^2 n_0 k/\hbar \ll Q$; being the length $l_{min}\sim{n_0}^{-1/2}$, the interparticle distance in this two-dimensional model. The resulting dispersion relation
\begin{eqnarray}\label{spinwavedispersion}
  \omega^2&=&\left(\omega_c+\frac{\hbar k^2}{2m}-\frac{4\pi\mu^2 n_0}{\hbar l_{min}}\right)^2\, ,
\end{eqnarray}
is a generalization for the dispersion of spin waves found in Refs.~\cite{Polyakov,misraMraklund}. The first term on the rhs of \eqref{spinwavedispersion} is the cyclotron frequency contributed by the external field, and the second one is the standard quantum contribution. It is the third term, originating in the  magnetic field induced by the spin self-interaction \eqref{eBf}, that is new. The new quantum correction (coming from multi-particle correlations) becomes comparable to the standard term when $(k l_{min})^2\sim\lambda_{classical} / l_{min}$, where $\lambda_{classical}$ is the classical electron radius.

\subsection{Plasma waves}

The dispersion for the ``plasma branch" (in  the  two-dimensional plasma) can be found using Eqs.~\eqref{per1}, \eqref{per2} and \eqref{per4},  \begin{equation}\label{}
  \omega^2=\frac{2\pi e^2 n_0 k}{m}+\omega_c^2+v_s^2 k^2+\frac{\hbar^2 k^4}{4 m^2}+\frac{\mu^2 n_0 \beta k^3}{m}\, .
\end{equation}
In this hybrid mode, the first term is the effective plasma frequency in a two-dimensional plasma \cite{krash}, the second term is the cyclotron contribution, and the third one comes from the thermal effects. The two last terms are the quantum contributions: the fourth term is the well known contribution of the Bohm potential, while the fifth one, proportional to $\beta\equiv\beta(k)$, is a totally  term originating in  contributions from the spin-spin interaction \eqref{47}.  Although the exact value of $\beta(k)$  requires numerical evaluation, approximate values  may be trivially  calculated for
small as well as large $k\, l_{min}$. Restricting to small  $k\, l_{min}$, we find $\beta(k)\sim 2\pi J_0(k l_{min})/k\, l_{min}\sim 2\pi/ k\, l_{min}$. The correlation term, then, goes as $k^2$ and competes with the thermal term in providing quadratic dispersion. Similar results have been found for plasmas with dipole-dipole interactions \cite{Andreev PRB 11}.

\section{Conclusions}
\label{sec7}

Starting from the $N$-particle Pauli-Scrodinger equation, we have derived the fluid equations for a spin quantum plasma, using a well-defined quantum mechanical averaging  procedure. This rather general averaging procedure, in addition to providing proper definitions for  macroscopic fluid variables,  allows us  to introduce inter particle corelations  caused by Coulomb as well as spin-spin interactions.

We would like to stress that the ability to express the interparticle interactions in terms of fluid variables is a fundamental strength of this class  of quantum averaging procedures. Consequently, in addition to the usual quantum forces due to the Bohm potential and the particle spin, new forces due to interparticle interactions appear; the latter can be viewed as effective electric and effective magnetic fields. The structural similarity of the new forces with the electromagnetic ones,  makes it possible to cast the whole quantum plasma dynamics  into a generalized vortex dynamics system with a conserved generalized helicity that has additional content owing to interparticle corelations.

As an application of the formalism, we calculated  the effects of  interparticle interactions on the propagation of electrostatic waves; the dispersion characteristics of both the spin and the plasma-cyclotron mode can be profoundly changed.

Finally, we want to emphasize that the  operational domain of the very straightforward quantum averaging procedure presented in this work, can be readily extended  to deal with various possible particle interactions that can take place in a quantum plasma.

\appendix

\section{Useful relations}
\label{apendiceA}

The following relations are useful in previous derivations
\begin{eqnarray}\label{ap1}
&&\partial_j\left(n\partial^j S^k\partial^i S^k \right)+\frac{n}{2}\partial^i\left(\partial_j S^k\partial_j S^k\right)\nonumber\\
&&+nS^k\partial^i\left(\frac{1}{n}\partial_j\left(n\partial^j S^k\right)\right)=n\partial^i\left(\frac{1}{n}\partial_k\left(n S^j\partial^k S^j\right)\right)\, ,\nonumber\\
&&
\end{eqnarray}
\begin{equation}\label{ap2}
  \varepsilon^{ijk}S_j\partial_m\left(n\partial_m S_k\right)=\partial_m\left(\varepsilon^{ijk}n S_j\partial_m S_k\right)\, .
\end{equation}

\section{Green functions}
\label{apendiceB}

The following integral
\begin{equation}\label{}
  \int d\mathbf r' G(\mathbf r',\mathbf r) e^{i \mathbf k\cdot(\mathbf r'-\mathbf r)}=\int d\vect\xi \frac{e^{i \mathbf k\cdot\vect\xi}}{\xi}\, ,
\end{equation}
with $\vect \xi=\mathbf r'-\mathbf r$ and $\xi=|\vect \xi|$, is readily evaluated for particles constrained to a two-dimensional plane,
\begin{equation}\label{}
  \int d\mathbf r' G(\mathbf r',\mathbf r) e^{i \mathbf k\cdot(\mathbf r'-\mathbf r)}=\int d\varphi d\xi e^{i k\xi\cos\varphi}=\frac{2\pi}{k}\, .
\end{equation}

Using the first part of \eqref{greenss2}, the integral
\begin{equation}\label{}
  \int d\mathbf r' G_{ij}(\mathbf r',\mathbf r) e^{i \mathbf k\cdot(\mathbf r'-\mathbf r)}=\int d\vect\xi\, G_{im}(\vect\xi)e^{i\mathbf k\cdot\vect \xi}\, ,
\end{equation}
is simply
\begin{eqnarray}\label{b4green2spin}
  \int d\mathbf r' G_{ij}(\mathbf r',\mathbf r) e^{i \mathbf k\cdot(\mathbf r'-\mathbf r)}&=&\int d\vect\xi\, \left[\left(\partial_i\partial_j-\delta_{ij}\nabla^2\right)\frac{1}{\xi} \right]e^{i\mathbf k\cdot\vect \xi}\nonumber\\
  &=&\frac{2\pi}{k}(\delta_{ij}k^2-k_i k_j)\, .
\end{eqnarray}

%%%%%%%%%%%%%%%%%%%%%%%%%%%%%%%%%%%%%%%%%%%%%

 \begin{acknowledgements}
 FAA thanks the CONICyT for a Becas Chile Postdoctoral Fellowship No. 74110049.
 \end{acknowledgements}


\begin{thebibliography}{17}

\bibitem{Haas PRE 00} F. Haas, G. Manfredi, M. Feix, Phys. Rev. E \textbf{62},
2763(2000).

\bibitem{Haasbook} F. Haas, {\it Quantum Plasmas: An Hydrodynamical Approach} (Springer Series on Atomic, Optical and Plasma Physics 65).

\bibitem{Marklund PRL 07} M. Marklund and G. Brodin, Phys. Rev. Lett. \textbf{98}, 025001 (2007).

\bibitem{pavel} P. A. Andreev and L. S. Kuz'menkov, Russian Phys. Jour. {\bf 50}, 1251 (2007).

\bibitem{brodinMHD} G. Brodin and M. Marklund, New J. Phys. 9, 277 (2007).

\bibitem{misracircularly} A. P. Misra, G. Brodin, M. Marklund and P. K. Shukla, J. Plasma Physics {\bf 76}, 857 (2010).
\bibitem{shuklawave} P. K. Shukla, Phys. Lett. A {\bf 369}, 312 (2007).
\bibitem{masood} W. Masood and A. Mushtaq, Phys. Lett. A {\bf 372}, 4283 (2008).
\bibitem{zamanian} G. Brodin, M. Marklund, J. Zamanian and M. Stefan, Plasma Phys. Control. Fusion {\bf 53}, 074013 (2011).
\bibitem{zama2} J. Zamanian, G. Brodin and M. Marklund, New Jour. Phys. {\bf 11}, 073017  (2009).
\bibitem{marksoliton} M. Marklund, B. Eliasson, and P. K. Shukla, Phys. Rev. E {\bf 76}, 067401 (2007).
\bibitem{misrasoliton} A. P. Misra and N. K. Ghosh, Phys. Lett. A {\bf 372}, 6412 (2008).
\bibitem{misra2soliton} A. P. Misra, N. K. Ghosh and P. K. Shukla, Phys. Plasmas {\bf 16}, 102309 (2009).
\bibitem{brodinpondero} G. Brodin, A. P. Misra, and M. Marklund, Phys. Rev. Lett. {\bf 105}, 105004 (2010).
\bibitem{misrawis} A. P. Misra, G. Brodin, M. Marklund, and P. K. Shukla, Phys. Rev. E {\bf 82}, 056406 (2010).
\bibitem{lundin} J. Lundin and G. Brodin, Phys. Rev. E {\bf 82}, 056407 (2010).
\bibitem{markferro} G. Brodin and M. Marklund, Phys. Rev. E {\bf 76}, 055403 (2007).
\bibitem{misramodu} A. P. Misra and P. K. Shukla, Phys. Plasmas {\bf 15}, 052105 (2008).
\bibitem{dae} D. -H. Ki and Y. -D. Jung, Appl. Phys. Lett. 99, 121506 (2011).
\bibitem{akbari} M. Akbari-Moghanjoughi, IEEE Trans. Plasma Sci. {\bf 40} 1330 (2012).
\bibitem{vitaly} V. Bychkov, M. Modestov, and M. Marklund, Phys. Plasmas {\bf 17}, 112107 (2010).
\bibitem{braun} S. Braun, F. A. Asenjo, and S. M. Mahajan, Phys. Rev. Lett. {\bf 109}, 175003 (2012).


\bibitem{Mahajan PRL 11} S. M. Mahajan and F. A. Asenjo, Phys. Rev. Lett. \textbf{107},
195003 (2011).

\bibitem{asenjo} F. A. Asenjo, V. Mu\~noz, J. A. Valdivia and S. M. Mahajan, Phys. Plasmas {\bf 18}, 012107 (2011).

\bibitem{jensC} J. Zamanian, M. Marklund and G. Brodin, arXiv:1303.3389 (2013).

\bibitem{MaksimovTMP 1999} L. S. Kuz'menkov and S. G. Maksimov,  Teor. i Mat. Fiz.,
 \textbf{118} 287 (1999) [Theoretical and Mathematical Physics \textbf{118} 227 (1999)].

\bibitem{Andreev PRB 11} P. A. Andreev, L. S. Kuz'menkov, M. I. Trukhanova, Phys. Rev. B \textbf{84}, 245401 (2011).

\bibitem{MaksimovTMP 2001} L. S. Kuz'menkov, S. G. Maksimov, and V. V. Fedoseev, Theor.
Math. Fiz. \textbf{126} 136 (2001) [Theoretical and Mathematical
Physics, \textbf{126} 110 (2001)].

\bibitem{Andreev IJMP 12} P. A. Andreev, L.S. Kuz'menkov,  Int. J. Mod. Phys. B \textbf{26} 1250186 (2012).

\bibitem{Landau Vol 3} L. D. Landau, E. M. Lifshitz, \emph{Quantum Mechanics: Non-Relativistic Theory}. Vol. 3 (3rd ed.). Pergamon Press,  (1977).

\bibitem{Weinberg book} S. Weinberg, \emph{Gravitation and Cosmology} (John Wiley and Sons, Inc., New York, 1972).

\bibitem{Klimontovich book} Yu. L. Klimontovich, \emph{Statistical Physics} [in Russian], Nauka, Moscow (1982); English transl., Harwood, New
York (1986).

\bibitem{MaksimovTMP 2001 b} L. S. Kuz'menkov, S. G. Maksimov, and V. V. Fedoseev, Theor.
Math. Fiz. \textbf{126} 258 (2001) [Theoretical and Mathematical
Physics, \textbf{126} 212 (2001)].


\bibitem{refvort} A more compact expression for quantum vorticity, in terms of a Clebsch potential, can be found in Ref.~\cite{Mahajan PRL 11}.

\bibitem{pavelproc} P. A. Andreev and L. S. Kuzmenkov, PIERS Proceedings, Moscow, Russia, August 19-23, 1055 (2012).

\bibitem{krash} M. V. Krasheninnikov and A. V. Chaplik, Sov. Phys. JETP {\bf 52}, 279 (1980).

\bibitem{Holland} P. R. Holland, {\it The quantum theory of motion} (Cambridge University Press, 1993).

\bibitem{Polyakov} P. A. Polyakov, Soviet Physics Jour. {\bf 22}, 310 (1979).

\bibitem{misraMraklund} A. P. Misra, G. Brodin, M. Marklund an P. K. Shukla, J. Plasma Phys. {\bf 76}, 857 (2010).


\end{thebibliography}
\end{document}